\documentclass[aps,pre,reprint,superscriptaddress,amsmath,amssymb,floatfix]{revtex4-2}
\usepackage{graphicx}
\usepackage{dcolumn}
\usepackage{bm}
\usepackage[colorlinks=true,
            linkcolor=blue,
            citecolor=blue,
            urlcolor=blue]{hyperref}
\begin{document}

\title{Opinion Formation in a Spatially Constrained Coevolving Nonlinear Voter Model}

\author{Phil Gwon Kim}
\thanks{These authors contributed equally to this work}
\affiliation{Department of Physics, KAIST, Daejeon 34141, Korea}

\author{Jaeyong Bae}
\thanks{These authors contributed equally to this work}
\affiliation{Department of Physics, KAIST, Daejeon 34141, Korea}

\author{Jaeseok Hur}
\affiliation{Department of Physics, KAIST, Daejeon 34141, Korea}

\author{Hawoong Jeong}
\email{hjeong@kaist.edu}
\affiliation{Department of Physics, KAIST, Daejeon 34141, Korea}
\affiliation{Center for Complex Systems, KAIST, Daejeon 34141, Korea}

\date{\today}

\begin{abstract}
We investigate a spatially constrained coevolving nonlinear voter model. 
Using a random geometric graph, we constrain the interaction range of voter dynamics.
If local rewiring is not possible, the discordant link is deleted.
Our results reveal absorbing states that differ not only in magnetization and activity, but also in mean degree and spatial state organization. 
By exploring dynamical and structural observables, we found that distinct regimes from the existing coevolving nonlinear voter model are characterized by a reduced consensus region, spatially segregated fragmentation, and isolated node formation. Additionally, we develop a phenomenological description of the evolution of the mean degree and terminal non-conserved quantities that is consistent with the numerical results. Our model highlights how local geometric accessibility reshapes the structure of the absorbing states.
\end{abstract}
                              
\maketitle

\section{Introduction}
Opinion dynamics and social networks often coevolve. Individuals may change their states under social influence, but they may also change their social contacts to avoid persistent disagreement~\cite{johnDontYouAny2015,paikDefriendingPolarizedAge2023}. 
Adaptive voter models formalize this feedback by treating individual state changes and the temporal evolution of the social network as coupled processes~\cite{gilCoevolutionAgentsNetworks2006,castellanoStatisticalPhysicsSocial2009b,grossAdaptiveCoevolutionaryNetworks2007,durrettGraphFissionEvolving2012,kozmaConsensusFormationAdaptive2008,starniniOpinionDynamicsStatistical2026}. 
One realization of this idea is the coevolving nonlinear voter model (CNVM). 
In this model, each node has one of two possible states. 
When an update occurs, where the update probability is nonlinear to the local disagreement, the discordant relation can be resolved either by a state change or by rewiring the discordant link.
The CNVM therefore provides a compact framework for studying how nonlinear social response and adaptive network restructuring jointly shape consensus, fragmentation, and active states~\cite{holmeNonequilibriumPhaseTransition2006,vazquezGenericAbsorbingTransition2008,castellanoNonlinearqvoterModel2009,minFragmentationTransitionsCoevolving2017,raduchaCoevolvingNonlinearVoter2018,demirelMomentclosureApproximationsDiscrete2014,vazquezTimescaleCompetitionLeading2007a,nardiniWhosTalkingFirst2008}.

However, real-world social interactions are often spatially constrained, so rewiring occurs by local proximity rather than being globally unconstrained, as commonly assumed in previous studies. In spatially embedded networks, rewiring is determined not only by state similarity but also by geographic accessibility~\cite{barthelemySpatialNetworks2011,mcphersonBirdsFeatherHomophily2001,onnelaGeographicConstraintsSocial2011.4.5.}. 
If an individual cuts a discordant link but cannot find a suitable target in the same state within the locally reachable range, rewiring cannot be completed, and the link is lost rather than replaced. 
Indeed, spatial correlations in voting patterns~\cite{borghesi2010spatial,fernandez2014voter,brownMeasurementPartisanSorting2021} and geographically grounded adaptive voter modeling~\cite{chuMicrodynamicsSpatialPolarization2021a} suggest that geographical proximity significantly influences opinion formation.

Therefore, we investigate opinion formation from both topological and geographical perspectives by incorporating a random geometric graph (RGG) into the CNVM. 
Our results show that both non-spatial and spatial fragmentation are possible, but that they are mainly determined by the nonlinearity of response. 
Furthermore, isolated nodes arise from rewiring failures caused by spatial constraints. 
In addition, we explain the empirical results using phenomenological approaches to the mean degree loss and other observables.

The remainder of this paper is organized as follows. 
In Sec.~\ref{sec2}, we introduce the CNVM on an RGG and define its dynamics and observables.
In Sec.~\ref{sec3}, we present the empirical results and characterize the resulting absorbing and finite-time dynamical regimes.
In Sec.~\ref{sec4}, we develop phenomenological descriptions of mean-degree relaxation and the terminal mean degree.
Finally, in Sec.~\ref{sec5}, we summarize our main findings and discuss their implications and possible future directions.

\section{\label{sec2} Model}
We study a coevolving nonlinear voter model on a spatially embedded network with local geometric constraints.
Spatial constraints are implemented by a random geometric graph (RGG), where interactions are allowed only between nodes separated by a distance smaller than $r$.
The same geometric constraint is imposed on rewiring, so the update rule differs from the conventional link-conserving CNVM.
We define the network, the dynamics, and the observables used below.

\subsection{\label{sec:model_RGG}Initial network setup}

The initial interaction network is an RGG with periodic boundary conditions~\cite{dallRandomGeometricGraphs2002}.
We place $N$ nodes independently and uniformly on the unit square $[0,1)^2$.
Two nodes $i$ and $j$ are connected if their periodic distance satisfies $d_{ij}<r$.
The interaction radius is chosen so that
$N r^2=8$,
giving the continuum estimate for the initial mean degree, 
\begin{equation}
k_0\simeq N\pi r^2=8\pi .
\end{equation}
For the system sizes used here, the initial networks are connected with high probability.
Mean degree conservation is not imposed during the subsequent dynamics.
Time is measured in Monte Carlo sweeps (MCS), where one MCS corresponds to $N$ microscopic update attempts.

\subsection{CNVM dynamics on the RGG}

Each node $i$ carries a binary state $s_i\in\{-1,+1\}$.
The initial states are assigned independently with equal probability, so the initial magnetization vanishes on average over ensembles.

Let $A_{ij}(t)$ denote the adjacency matrix at time $t$.
The degree of node $i$ is
\begin{equation}
k_i(t)=\sum_j A_{ij}(t),
\end{equation}
and the number of active links incident to node $i$ is
\begin{equation}
a_i(t)
=
\sum_j A_{ij}(t)\mathbf{1}\{s_i(t)\neq s_j(t)\}.
\end{equation}
An active link is a link connecting two nodes with different states.
For $k_i(t)>0$, the local active link density is
\begin{equation}
\rho_i(t)=\frac{a_i(t)}{k_i(t)}.
\end{equation}
If $k_i(t)=0$, the node is isolated, and no update occurs when it is selected.

At each microscopic update, a node $i$ is chosen uniformly at random.
If $k_i(t)>0$, it is activated with probability $\rho^q_i(t)$, where $q$ controls the nonlinear response to local disagreement~\cite{castellanoNonlinearqvoterModel2009, minFragmentationTransitionsCoevolving2017}.
If the node is activated, then with probability $1-p$ it flips,
\begin{equation}
s_i\to -s_i .
\end{equation}
With probability $p$, a rewiring event occurs.
One active link $(i,j)$ incident to $i$ is chosen uniformly and removed.
The node $i$ then searches for same-state candidates within the local interaction radius,
\begin{equation}
\mathcal{C}_i
=
\left\{
\ell\neq i:
s_\ell=s_i,\,
A_{i\ell}=0,\,
d_{i\ell}<r
\right\}.
\end{equation}

If $\mathcal{C}_i$ is nonempty, one node $\ell \in \mathcal{C}_i$ is chosen uniformly at random, and a new link $(i,\ell)$ is added. If $\mathcal{C}_i$ is empty, the removed link is not replaced. This conditional deletion is distinct from the explicit ``rewire-to-none'' rule studied in the coevolving voter model~\cite{kurehFittingBreakingNonlinear2020}. In our model, deletion does not occur as a separate update rule. Rather, it happens endogenously when the updating node has no geometrically accessible same-state candidate.

This local geometric restriction is the main departure from the standard CNVM. Because failed rewiring attempts remove links without replacement, the total number of links and the mean degree are no longer conserved, and isolated nodes can emerge during the dynamics.

\subsection{Dynamical and structural observables}

To quantify the state of the system, we use the following macroscopic observables.
From the adjacency matrix at time $t$, the total numbers of links and active links are
\begin{equation}
E(t)=\frac{1}{2}\sum_{i,j}A_{ij}(t), \quad E_{\rm active}(t)=\frac{1}{2}\sum_{i}a_i(t).
\end{equation}
The global active link density is defined as the ratio of active links to total links,
\begin{equation}
\rho(t)=\frac{E_{\rm active}(t)}{E(t)}.
\end{equation}
An absorbing state is defined as a configuration with $\rho(t)=0$. 
Global state ordering is measured by the absolute magnetization,
\begin{equation}
|m(t)| = \left| \frac{1}{N} \sum_{i=1}^{N}s_i(t) \right|.
\end{equation}
To quantify fragmentation, we measure the size of the largest connected component, $S_{\max}(t)$, and use the normalized fraction $S_{\max}(t)/N$.
The mean degree is given by
\begin{equation}
k(t)=\frac{1}{N}\sum_{i=1}^{N}k_i(t),
\end{equation}
and the number of isolated nodes is
\begin{equation}
N_{\rm iso}(t) = \sum_{i=1}^{N} \mathbf{1}\{k_i(t)=0\}.
\end{equation}

\begin{figure*}[ht]
    \centering
    \includegraphics[width=\linewidth]{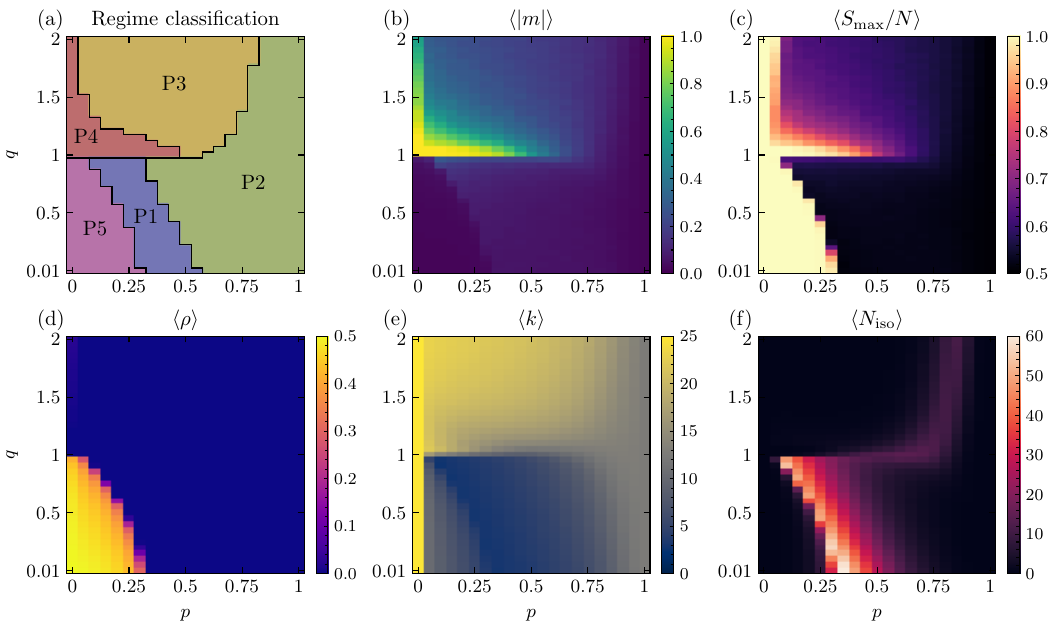}
    \caption{
    Summary of empirical regimes and macroscopic observables.
    (a) Empirical regime classification, (b) $\langle |m| \rangle$,
    (c) $\langle S_{\max}/N \rangle$, (d) $\langle \rho \rangle$,
    (e) $\langle k \rangle$, and (f) $\langle N_{\rm iso}\rangle$.
    Simulations are performed with $N=2^{12}$ and maximum observation time
    $T_{\max}=2^{15}$ Monte Carlo sweeps.
    Each parameter point $(p,q)$ is averaged over $2^{10}$ independent realizations.
    The rewiring probability is varied as $p=\{0,0.05,0.10,\ldots,1.00\}$,
    and the nonlinearity is scanned over $q=\{0.01,0.05,0.10,\ldots,2.00\}$.
    }
    \label{fig:summary_six_panel}
\end{figure*}

\section{\label{sec3} Empirical Results}

To summarize the parameter space, we construct an empirical regime map from macroscopic observables evaluated at the maximum observation time $T_{\max}$.
For each parameter pair $(p,q)$, we define the feature vector

\begin{equation}
\begin{aligned}
{\bf x}(p,q;T_{\max})=\big(&\langle |m| \rangle,
\langle \rho \rangle,
\langle S_{\max}/N \rangle,\langle k \rangle,
\langle N_{\rm iso}\rangle\big).
\end{aligned}
\label{eq:feature_vector}
\end{equation}
Here, $\langle \cdot \rangle$ denotes an ensemble average over independent realizations at fixed $(p,q)$, evaluated at the final observation time, $T_{\max}$.
Before clustering, each component of ${\bf x}$ is standardized over the parameter grid.
We then apply $k$-means clustering.
The number of clusters is selected using the silhouette score~\cite{rousseeuwSilhouettesGraphicalAid1987}, yielding five empirical regimes labeled P1--P5. Details of the clustering procedure and robustness checks are given in Appendix~\ref{app:k_means}.

Fig.~\ref{fig:summary_six_panel} shows the resulting empirical regime map and the observables used for classification.
The labels P1--P5 summarize patterns observed within the finite simulation window. Complementary results with different system size $N$ are given in Appendix~\ref{app:size_robustness}.

We first discuss the high-$|m|$ regime P4, then distinguish topological fragmentation from spatial segregation, relate isolated node formation to failed rewiring, and finally examine the low-$p$, low-$q$ regime that remains active up to the observation time.

\subsection{Finite-size effects on consensus regime}

Regime P4 has small activity and relatively large magnetization, and therefore appears to exhibit consensus in finite simulations.
This interpretation is finite-size sensitive.
In absorbing state systems, finite-size fluctuations can produce consensus-like absorbing outcomes that do not necessarily imply a thermodynamic consensus phase~\cite{vazquezGenericAbsorbingTransition2008, minFragmentationTransitionsCoevolving2017}.

\begin{figure}[t]
    \centering
    \includegraphics[width=\columnwidth]{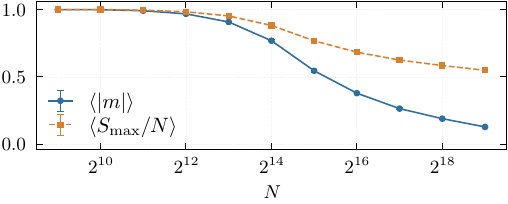}
    \caption{system size dependence at $p=0.25$ and $q=1$.
    These simulations use $T_{\max}=2^{16}$ MCS, and all runs shown reach absorption within this time.
    The absorbed-run magnetization decreases strongly with $N$, whereas $\langle S_{\max}/N\rangle$ remains larger, consistent with a dominant connected cluster with residual opposite-state clusters. Error bars represent the standard error.}
    \label{fig:p4_consensus_like}
\end{figure}

A robust consensus phase should remain stable as the system size increases.
Fig.~\ref{fig:p4_consensus_like} shows the system size dependence at a representative point in P4, $p=0.25$ and $q=1$.
All realizations in this test reach absorption within $T_{\max}=2^{16}$ MCS.
At the main system size used in Fig.~\ref{fig:summary_six_panel}, $N=2^{12}$, the absorbed state is almost fully ordered.
As $N$ increases, however, $\langle |m| \rangle_{\rm abs}$ decreases steadily.
The high-$|m|$ branch near $q\simeq1$ is therefore not stable under increasing system size.

This loss of global magnetization does not indicate homogeneous disorder.
In Fig.~\ref{fig:p4_consensus_like}, $\langle S_{\max}/N\rangle$ remains substantially larger than $\langle |m|\rangle$ over the same range of $N$.
This is consistent with a dominant connected cluster persisting while residual opposite-state clusters reduce the global magnetization.
Thus, with increasing $N$, the outcome is more naturally described as a fragmented absorbing configuration with a dominant state cluster. These results support the interpretation of P4 as a finite-size consensus-like absorbing regime rather than a robust thermodynamic consensus phase. A complementary long-time test of the sensitivity to finite rewiring is reported in Appendix~\ref{app:P4_finite_rewiring}.

\subsection{Spatially segregated fragmentation}

Initially, states are uncorrelated with position because states are assigned independently on the RGG.
During the CNVM dynamics, state flips and rewiring can produce topological fragmentation by eliminating active contacts between opposite state vertices.
Because the nodes are embedded in two-dimensional space, we ask whether this topological fragmentation is accompanied by real-space segregation of states.

To quantify real-space segregation, we define the set of geometrically neighboring pairs as
\begin{equation}
{\cal N}_{\rm sp}
=
\{(i,j):i<j,\ d_{ij}<r\},
\end{equation}
where $d_{ij}$ is the periodic distance and $r$ is the RGG interaction radius.

The same-state fraction among these spatial neighbors is
\begin{equation}
f_{\rm sp}
=
\frac{1}{|{\cal N}_{\rm sp}|}
\sum_{(i,j)\in{\cal N}_{\rm sp}} \delta_{s_i,s_j}.
\end{equation}

Since a global imbalance between $+1$ and $-1$ states can increase this fraction even in the absence of spatial structure, we subtract the random shuffle baseline at fixed state counts,
\begin{equation}
f_{\rm rand}
=
\frac{n_+(n_+-1)+n_-(n_--1)}{N(N-1)}.
\end{equation}
Here $n_+$ and $n_-$ are the total numbers of $+1$ and $-1$ states in the final state, respectively. 

Subtracting this baseline gives the spatial enrichment,
\begin{equation}
\Delta_{\rm sp}
=
f_{\rm sp}-f_{\rm rand}.
\end{equation}

\begin{figure}[t]
\centering
\includegraphics[width=\columnwidth]{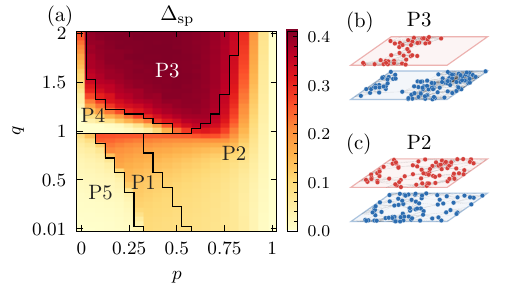}
\caption{
Spatial segregation in final configurations.
(a) Spatial enrichment $\Delta_{\rm sp}$ in the $(p,q)$ plane.
The labels P1--P5 denote the empirical regimes from Fig.~\ref{fig:summary_six_panel}.
(b,c) Representative absorbing configurations for P3 and P2, respectively. P3 illustrates spatially segregated fragmentation, whereas P2 illustrates mainly topological fragmentation with weak real-space segregation.}
\label{fig:spatial_segregation}
\end{figure}

Fig.~\ref{fig:spatial_segregation}(a) shows that P3 is distinguished from the other absorbing fragmented regimes by a strong enhancement of the spatial enrichment.
For instance, P2 can be both absorbing and fragmented in the graph-theoretic sense, but it exhibits only weak spatial enrichment.
Thus, P2 primarily represents topological fragmentation, in which discordant links are removed without significant real-space separation of the two state types.
In contrast, P3 corresponds to spatially segregated fragmentation, in which graph fragmentation is accompanied by real-space state-domain separation.

Standard link-based assortativity~\cite{newmanMixingPatternsNetworks2003} cannot distinguish these absorbing configurations, because all surviving links are same-state links when $\rho=0$. The spatial enrichment $\Delta_{\rm sp}$ is therefore needed to distinguish real-space segregation from graph fragmentation alone.

A plausible mechanism for this spatial segregation is the nonlinear activation rule. For $q>1$, updates are biased toward nodes with high local active link density $\rho_i$.
These nodes are typically located near interfaces between opposite state regions. When the rewiring probability is not too large, state flips can therefore move and smooth these interfaces before discordant links are fully eliminated. The subsequent removal of the remaining discordant links freezes the coarsened state domains into a spatially segregated, fragmented absorbing state.

\subsection{Cumulative failed rewiring and isolated node formation}

A distinctive feature of the present model is that link loss can accumulate during the dynamics.
Although each link-loss event occurs through failed rewiring, the isolated node pattern is not a monotonic function of the rewiring probability $p$.
This indicates that isolated node formation is controlled not by $p$ alone, but by the accumulation of failed rewiring events over the lifetime of the dynamics.
We therefore measure the cumulative number of rewiring attempts, $N_{\rm rew}$, and the cumulative number of failed rewiring events, $N_{\rm fail}$.

\begin{figure}[t]
    \centering
    \includegraphics[width=\linewidth]{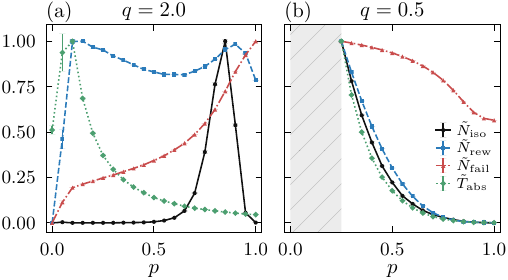}
    \caption{Normalized $p$-dependence of the observables for $q = 0.5$ and $q = 2.0$. The tilde denotes that each ensemble-averaged observable is normalized by its maximum value along the corresponding $q$-cut ($\tilde{O} \equiv \langle O \rangle / \langle O \rangle_{\max}$). As an example, the black curve for $\tilde{N}_{\rm iso}$ represents the normalized values derived from each corresponding $q$-cut in Fig.\ref{fig:summary_six_panel} (f). 
    (a) At $q = 2.0$, $\tilde{N}_{\rm fail}$ reaches its maximum at $p = 1$, whereas $\tilde{N}_{\rm iso}$ exhibits a distinct peak at $p < 1$.
    (b) At $q = 0.5$, $\tilde{N}_{\rm iso}$ closely follows the failed rewiring count $\tilde{N}_{\rm fail}$. The shaded region indicates the active regime where no runs reached the absorbing state within $T_{\max}$, and is thus excluded from the statistics. Error bars represent the standard error of the mean.}
    \label{fig:rewiring_tracking}
\end{figure}

Fig.~\ref{fig:rewiring_tracking} compares $N_{\rm iso}$, $N_{\rm rew}$, $N_{\rm fail}$, and absorption time $T_{\rm abs}$ for representative cuts $q=2.0$ and $q=0.5$.
To facilitate comparison across different scales, we define the normalized observable $\tilde{O} \equiv \langle O \rangle / \langle O \rangle_{\max}$, where each ensemble-averaged quantity is divided by its maximum value along the corresponding $q$-cut. 

For $q=2$, $\tilde{N}_{\rm fail}$ increases toward $p=1$, but $\tilde{N}_{\rm iso}$ peaks at a large value below $p=1$.
At $p=1$, state flips are absent, and the initial RGG already contains all geometric links within the interaction radius.
After an active link is removed, the updating node therefore has no same-state geometric non-neighbor available, so rewiring fails.
This fixed state pruning produces many failed rewiring events, but it does not keep regenerating active links around low-degree nodes.
Once the discordant links of such a node have been removed, its remaining same-state links are inactive and cannot be depleted further.
For $p<1$, rare state flips change this local balance.
They can reduce the failure rate of individual rewiring attempts by creating same-state geometric non-neighbors, but they can also turn the remaining links of low-degree nodes back into active links.
When that happens, the local active link density of a low-degree node can be close to one, so $\rho^q_i(t)$ remains large even for $q>1$.
At large $p<1$, rewiring is still frequent enough to remove these reactivated links before the local configuration relaxes.
This provides a mechanism for why the isolated node count peaks below $p=1$, whereas the failed rewiring count is maximal at $p=1$.
The low-degree tail in Appendix~\ref{app:degree_k} is consistent with this interpretation.

For $q=0.5$, the peak of $\tilde{N}_{\rm iso}$ closely follows that of $\tilde{N}_{\rm fail}$.
In this regime, $\rho_i^q(t)$ remains sizable even when $\rho_i(t)$ is small, so weakly discordant nodes can continue to activate.
The resulting active lifetime is long, so the accumulated number of rewiring opportunities, rather than the instantaneous failure probability, becomes the dominant factor.
Thus, isolated-node formation is controlled primarily by accumulated failed rewiring, rather than by the instantaneous value of the rewiring probability alone.
Consistently, for $q=0.5$, $\tilde{N}_{\rm iso}$ is largest in the region where both $\tilde{N}_{\rm fail}$ and $\tilde{T}_{\rm abs}$ are large.

\subsection{Finite-time active regime at low \texorpdfstring{$p$}{p} and low \texorpdfstring{$q$}{q}}

P5 corresponds to the low-$p$, low-$q$ region where a substantial fraction of runs remain active throughout the finite observation window used in the simulations.
Thus, P5 should be interpreted as a finite-time active regime, not as an established asymptotic active phase.

To test whether this activity persists, we performed longer simulations at a representative point, $p=0.1, q=0.1$, for system sizes up to $N=2^{13}$ (See Appendix~\ref{app:abs_time}). 
At this representative point, all runs reached an absorbing state within the extended simulation time.
The mean absorption time remains approximately of the same order in Monte Carlo sweep units over the accessible system sizes, corresponding to roughly linear scaling in microscopic update units $\tau$.
This behavior contrasts with the standard degree-preserving CNVM, where the lifetime of the finite-size active regime can grow exponentially with system size, $\tau\sim\exp(N)$~\cite{minFragmentationTransitionsCoevolving2017}.
The comparison suggests that spatially constrained rewiring may accelerate absorption by promoting local topological fragmentation rather than global mixing.
A finite-time scaling analysis of the P5--P1 boundary is given in Appendix~\ref{app:eff_fss}.

\section{\label{sec4} mean degree dynamics and absorption}
The previous section characterized the empirical regimes at a fixed observation time. We now focus on a dynamical feature specific to the present model, the irreversible loss of links caused by failed rewiring attempts. Because the mean degree changes only when a rewiring attempt fails, its evolution quantifies the coupling between nonlinear activation, local spatial constraints, and absorption.

We first compare the relaxation of the active link density and the mean degree, and use this ordering to motivate a coarse-grained description of mean degree dynamics. We then treat the terminal mean degree as an absorption-conditioned observable. This yields a finite-time stopping description in which the absorption time and terminal degree are evaluated within an effective structure.

\subsection{\label{sec:degree_relax}Time evolution analysis of mean degree loss with relative relaxation}

We now describe the relaxation of the mean degree from a phenomenological point of view. 
Our goal is to construct the minimal coarse-grained description of the ensemble-averaged mean degree trajectory that is consistent with the observed relative relaxation.

In the present model, the mean degree decreases only when a rewiring attempt fails.
Such an event requires an active update, a rewiring attempt with probability $p$, and the absence of an available same-state geometric target.

At $t=0$, the RGG already contains all geometric links within the interaction radius, so an updating node typically has no available same-state non-neighbor inside its local neighborhood.
Rewiring failures dominate at early times, and the mean degree can drop sharply once active links are selected for rewiring. As link losses and state flips reshape the local neighborhoods, available same-state targets can appear, while the active interface gradually shrinks. The resulting loss rate is therefore not constant, but decreases with time in an effective coarse-grained sense.

We model this slowing down phenomenologically. At fixed $(p,q)$, let $K(t)\equiv \langle k(t)\rangle$ and define the excess mean degree $\tilde K(t)=K(t)-K_s$, where $K_s$ is the reference endpoint used in the fit. We assume an aging loss rate,
\begin{equation}
\frac{d\tilde K(t)}{dt}
=
-\Lambda_{p,q}(t)\tilde K(t),
\quad
\Lambda_{p,q}(t)=\Lambda^0_{p,q}(t+t_0)^{-\gamma_{p,q}} .
\label{eq:degree_phenom_rate}
\end{equation}

The rate $\Lambda_{p,q}(t)$ is an effective loss rate for the absorbed run ensemble average.
The cutoff $t_0$ absorbs the short initial transient, while $0<\gamma_{p,q}<1$ sets the strength of the slowing down.
In the fits below, $K_s$ is fixed from the observed endpoint of the common pre-absorption window,
\begin{equation}
K_s = K(T_{\rm abs}^{\min}),
\qquad
T_{\rm abs}^{\min}=\min_a T_{\rm abs}^{(a)} ,
\end{equation}
rather than predicted from the phenomenological relaxation equation.

For $t$ beyond the short initial cutoff, integrating this aging rate equation gives the stretched exponential function
\begin{equation}
\tilde K_{\rm fit}(t)
=
\tilde K(0)
\exp\left[
-\left(\frac{t}{\tau}\right)^\beta
\right],
\quad
\beta=1-\gamma_{p,q}.
\label{eq:single_k_fit}
\end{equation}

Thus, the stretched exponential is used here as a phenomenological consequence of an aging loss rate, not as a microscopic mean-field prediction.

We next ask whether a single time scale component is sufficient.
For this purpose, we compare the relaxation of the active link density and the mean degree in absorbed trajectories.
For each absorbed trajectory $a$, with absorption time $T_{\rm abs}^{(a)}$, we normalize the two observables over the pre-absorption interval $0\le t\le T_{\rm abs}^{(a)}$ as
\begin{equation}
\rho_f^{(a)}(t)=\frac{\rho_a(t)}{\rho_a(0)},
\qquad
k_f^{(a)}(t)=
\frac{k_a(t)-k_a(T_{\rm abs}^{(a)})}
{k_a(0)-k_a(T_{\rm abs}^{(a)})}.
\end{equation}
Because this normalization requires nonzero degree loss, the $p=0$ line is excluded from this analysis. We defined the relative relaxation for each run as follows
\begin{equation}
D^{(a)}
=
\frac{1}{T_{\rm abs}^{(a)}}
\int_0^{T_{\rm abs}^{(a)}}
\left[
\rho_f^{(a)}(t)-k_f^{(a)}(t)
\right]dt.
\label{eq:D}
\end{equation}

If $D^{(a)}>0$, the normalized mean degree decreases earlier than the activity, which we call topology-first relaxation.
If $D^{(a)}<0$, the activity decreases earlier than the mean degree, which we call activity-first relaxation. 

\begin{figure}[t]
\centering
\includegraphics[width=\columnwidth]{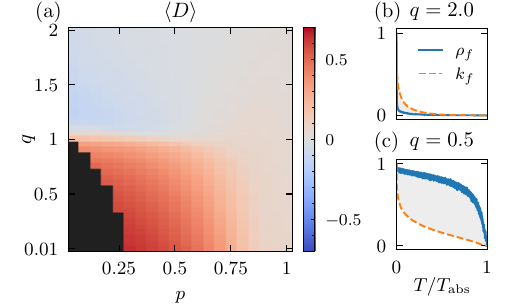}
\caption{
Ensemble-averaged relative relaxation $\langle D\rangle$ in Eq.~\eqref{eq:D} for absorbed trajectories.
(a) $\langle D\rangle$ in the $(p,q)$ plane. The $p=0$ line is excluded because the mean degree is conserved and the normalization of $k_f$ is undefined. 
Black cells indicate that no runs reached absorption within the observation time.
Representative normalized trajectories at $p=0.25$ for $q=2.0$ (b) and $q=0.5$ (c).
Positive $D$ corresponds to topology-first relaxation; negative $D$ corresponds to activity-first relaxation.
}
\label{fig:relaxation_ordering}
\end{figure}

Fig.~\ref{fig:relaxation_ordering} shows the ensemble-averaged relative relaxation $\langle D\rangle$ across the parameter space and presents representative normalized trajectories for the two relaxation routes. 
Panel~(a) reveals that the relaxation route changes systematically with $q$.

For $q>1$, $\langle D\rangle$ is close to zero or negative over a broad region, indicating that activity relaxation and degree loss are not resolved as two widely separated macroscopic stages.
We therefore use the single component closure in Eq.~\eqref{eq:single_k_fit} as the minimal description for this regime. 

For $q<1$, most absorbed parameter cells have $\langle D\rangle>0$, indicating topology-first relaxation.
This behavior reflects the interplay between residual activity and the local reservoir of available same-state rewiring targets.
Immediately after the initial transient, activity is still high, but the same-state target reservoir is small because the initial RGG already contains all local geometric links.
Rewiring attempts therefore fail frequently, producing a rapid drop in the mean degree.
At later times, state flips and previous link losses partially reshape the local neighborhoods, so successful rewiring becomes more possible, and the degree-loss rate decreases.
However, for $q<1$, weakly discordant nodes can still activate for a long time because $\rho_i^q$ remains sizable even when $\rho_i$ is small.
The mean degree therefore develops a slow erosion tail sustained by this long-lived residual activity.
This empirical temporal separation motivates a two-component closure,

\begin{align}
\tilde{K}_{\rm fit}(t)
= \tilde{K}(0)
\bigg[
&w\exp\left[-\left(\frac{t}{\tau_1}\right)^{\beta_1}\right] \notag\\
&+ (1-w)\exp\left[-\left(\frac{t}{\tau_2}\right)^{\beta_2}\right]
\bigg].
\label{eq:two_k_fit}
\end{align}
with $\tau_1 < \tau _2$.
Eq.~\eqref{eq:two_k_fit} is therefore used as an effective time scale decomposition. 
The fast component captures failure-dominated degree loss after the initial transient, when same-state target reservoirs are still scarce, while the slow component captures residual erosion sustained by long-lived activity after local neighborhoods have partially relaxed.

\begin{figure}[t]
\centering
\includegraphics[width=\columnwidth]{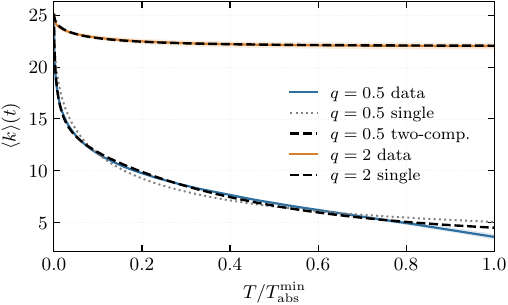}
\caption{
Stretched exponential fits to the mean degree.
Representative absorbed-run ensemble averages of $K(t)=\langle k(t)\rangle$ are shown at $p=0.25$ for $q=0.5$ and $q=2.0$.
The time axis is normalized by $T_{\rm abs}^{\min}$, and only the common pre-absorption window $0\le t/T_{\rm abs}^{\min}\le1$ is used.
For $q=0.5$, the gray dotted curve shows the single stretched exponential fit and the black dashed curve shows the two-component fit.
For $q=2.0$, the black dashed curve shows the single stretched exponential fit. Error bands represent the standard deviation over runs.}
\label{fig:relaxation_fit}
\end{figure}

Fig.~\ref{fig:relaxation_fit} shows the resulting fits for representative parameter values at $p=0.25$.
For $q=0.5$, the single stretched exponential exhibits a systematic deviation over the early and intermediate relaxation window, whereas the two-component form follows the trajectory over most of the common pre-absorption interval.
For $q=2.0$, the single stretched exponential already captures the dominant relaxation envelope, consistent with the absence of a clearly separated topology-first stage.
These fits should therefore be interpreted as phenomenological closures motivated by the relative relaxation.
In particular, $K_s$ is fixed by the observed finite-time endpoint of the fitting window.
The terminal mean degree and the absorption time are treated separately below within a finite-time stopping process description.

\subsection{\label{sec:absorption_model}Absorption as a stopping process}

We next consider the terminal properties of the absorbing state. In the present model, absorption is defined by the disappearance of active links, $\rho=0$. This condition, however, does not uniquely determine the final graph structure. The same absorbing condition can be reached after different amounts of link loss, depending on how nonlinear activation and failed rewiring events are organized along the trajectory. The terminal state must therefore be characterized by both the absorption time and the mean degree remaining when the dynamics stop.

We describe these quantities using a reduced stopping process ansatz in which absorption is treated as a first-passage event through one of several effective routes. Each route represents a trajectory class specified by a route weight, a first-passage law, and a terminal degree function. Because the microscopic state space of the adaptive graph is large, this route description provides a low-dimensional phenomenological closure for the finite-time absorption probability, the conditional absorption time, and the conditional terminal mean degree. Such first-passage event based coarse-grained descriptions have been used to replace unresolved microscopic dynamics by a small number of effective completion paths fitted to observed passage-time statistics~\cite{rivera2021inferring}.

The multiple route representation accounts for the coexistence of distinct absorption histories in the $(p,q)$ plane. For $q<1$, weak local discordance can still generate updates, so that repeated failed rewiring attempts may erode the graph before activity vanishes. For $q>1$, activation is suppressed unless the local discordance is sufficiently large, and absorption can occur with comparatively little degree loss. The route representation separates these degree-eroding and degree-preserving histories at the coarse-grained level.

Each route $i$ is specified by three quantities. The route weight $\alpha_i(p,q)$ is the mixture weight of route $i$ at the parameter point $(p,q)$. The cumulative first-passage law
\begin{equation}
F_i(T;p,q)=\Pr(T_i\le T)
\label{eq:route_cdf_main}
\end{equation}
gives the probability that absorption has occurred by time $T$, conditional on route $i$. The terminal degree function $d_i(p)$ assigns the mean degree associated with absorption through that route. Thus $F_i$ controls the absorption time distribution, whereas $d_i$ sets the terminal mean degree assigned to that route.

We use three effective routes, denoted by $L$, $M$, and $H$. Their terminal degree functions represent a degree-eroding branch, an intermediate branch, and a degree-preserving branch. The route labels refer to terminal degree, while the associated first-passage laws describe the absorption time statistics. The terminal degree functions are constrained by three degree levels: the fragmented residual level $k_{\min}$, the reference level of discordant link removal $k_{\rm half}=k_0/2$, and the initial degree level $k_0$. The level $k_{\rm half}$ follows from the unbiased initial state assignment. On average, roughly half of a node's initial neighbors have the opposite state, so removing discordant links gives a reference scale near $k_0/2$.

The endpoint functions are chosen to be monotone in $p$. The degree-eroding branch increases from $k_{\min}$ toward $k_{\rm half}$ as rewiring becomes more frequent, whereas the degree-preserving branch decreases from $k_0$ toward $k_{\rm half}$ as discordant links are progressively removed. We therefore use
\begin{equation}
\begin{aligned}
d_L(p)
&=
k_{\min}
+
(k_{\rm half}-k_{\min})p^{\nu_L},\\
d_H(p)
&=
k_{\rm half}
+
(k_0-k_{\rm half})(1-p)^{\nu_H},\\
d_M(p)
&=
\omega_M d_L(p)+(1-\omega_M)d_H(p),
\end{aligned}
\label{eq:main_degree_functions}
\end{equation}
with $\nu_L,\nu_H>0$ and $0\le\omega_M\le1$. 
The empirical construction of these endpoint functions is given in Appendix~\ref{app:absorbing_closure}.

The route ansatz is summarized as
\begin{equation}
\begin{aligned}
{\cal R}_L &: \{\alpha_L(p,q),F_L(T;p,q),d_L(p)\},\\
{\cal R}_M &: \{\alpha_M(p,q),F_M(T;p,q),d_M(p)\},\\
{\cal R}_H &: \{\alpha_H(p,q),F_H(T;p,q),d_H(p)\}.
\end{aligned}
\label{eq:route_ansatz_summary}
\end{equation}
The weights specify the mixture of routes, the first-passage laws specify the absorption time statistics, and the terminal degree functions specify the route-level terminal degrees.

For a route dependent quantity $\phi_i(t)$, define the finite window route moment
\begin{equation}
{\cal M}_i[\phi](p,q)
=
\int_0^{T_{\max}}
\phi_i(t)\,\partial_tF_i(t;p,q)\,dt .
\label{eq:main_route_moment}
\end{equation}
Here $\partial_tF_i(t;p,q)\,dt$ is the probability that absorption through route $i$ occurs in the interval $[t,t+dt)$.

The probability of absorption within the finite observation window is
\begin{equation}
{\cal Z}(p,q)
=
\sum_{i\in\{L,M,H\}}
\alpha_i(p,q){\cal M}_i[1](p,q)
\label{eq:main_Z}
\end{equation}
where ${\cal M}_i[1](p,q)=F_i(T_{\max};p,q)$.
This quantity is the finite window absorption probability and the normalizing denominator for absorption conditioned averages.
Taking $\phi_i(t)=t$ gives the conditional mean absorption time,
\begin{equation}
\tau_{\rm abs}(p,q)
=
\frac{
\sum_{i\in\{L,M,H\}}
\alpha_i(p,q){\cal M}_i[t](p,q)
}{
{\cal Z}(p,q)
}.
\label{eq:main_tau_abs}
\end{equation}
The terminal mean degree is obtained by taking $\phi_i(t)=d_i(p)$,
\begin{equation}
k_{\rm abs}(p,q)
=
\frac{
\sum_{i\in\{L,M,H\}}
\alpha_i(p,q){\cal M}_i[d_i](p,q)
}{
{\cal Z}(p,q)
}.
\label{eq:main_k_abs}
\end{equation}

\begin{figure}[t]
\centering
\includegraphics[width=\columnwidth]{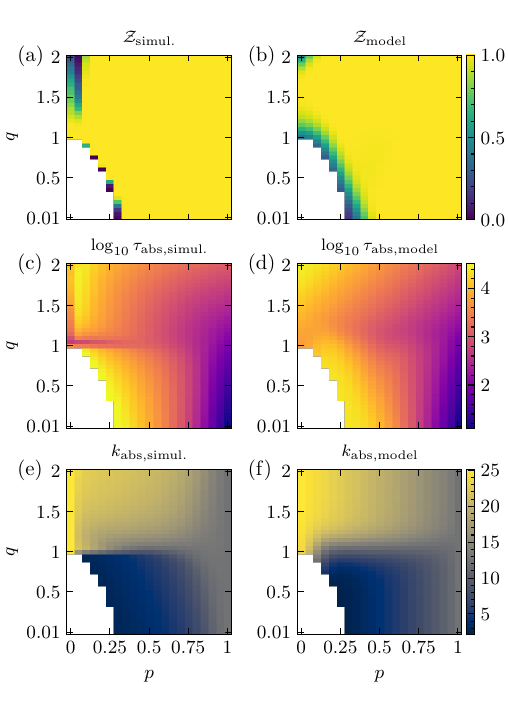}
\caption{
Comparison between simulations and the phase-type stopping model.
The panels show the finite-time absorption probability, the conditional mean absorption time, and the conditional terminal mean degree in the $(p,q)$ plane.
Simulation results (a)-(c)-(e) are compared with the model results (b)-(d)-(f) of ${\cal Z}(p,q)$, $\log \tau_{\rm abs}(p,q)$, and $k_{\rm abs}(p,q)$.
Blank regions indicate unabsorbed states, corresponding to the nonzero region in Fig.~\ref{fig:summary_six_panel}(d).
The same route weights and first-passage rates enter all three observables, while the terminal degree functions $d_i(p)$ determine the degree assigned to each absorbed route.}
\label{fig:phase_type_comparison}
\end{figure}

Equations~\eqref{eq:main_Z}--\eqref{eq:main_k_abs} define the reduced stopping closure. We fit the route weights, the route-dependent first-passage rates, and the terminal degree functions jointly, and compare the resulting predictions with the three measured observables in Fig.~\ref{fig:phase_type_comparison}.

Fig.~\ref{fig:phase_type_comparison} shows that the reduced route description captures the main trends of all three terminal observables. The comparison is constrained because all three observables use the same route structure. The same weights and first-passage rates determine the finite-time absorption probability and the absorption time statistics, and the same absorbed routes are used to assign the terminal degree. The result summarizes the competition among degree-eroding, mixed, and degree-preserving terminal branches.

\section{\label{sec5} Conclusion}
We investigated the role of spatially constrained interactions in the coevolving nonlinear voter model by introducing a random geometric graph. Our framework makes it possible to analyze the spatial fragmentation of opinions in addition to topological fragmentation. We first examined the macroscopic observables in the $(p,q)$ plane, including the magnetization $\langle{|m|}\rangle$, active link density $\langle{\rho}\rangle$, largest component fraction $\langle{S_{\max}/N}\rangle$, mean degree $\langle{k}\rangle$, and number of isolated nodes $\langle{N_{\rm iso}}\rangle$. These results reveal that macroscopic properties of the state configuration cannot be fully characterized by $\langle{|m|}\rangle$ and $\langle{\rho}\rangle$ alone.

Our model is distinct from the previous CNVM~\cite{minFragmentationTransitionsCoevolving2017} in that the global consensus region reduces when $q>1$ and high magnetization exhibits a finite-size effect. This highlights the difficulty of achieving global consensus through purely local interactions.
The fragmented absorbing states can be classified into two regimes: topological fragmentation and spatially segregated fragmentation. These two regimes are mainly determined by the nonlinearity $q$, and the spatial enrichment $\Delta_{\rm sp}$ characterizes spatial segregation of the state configurations.

A core mechanism driving these regimes is the accumulation of rewiring failures. 
During the initial stages of the dynamics, fluctuations can induce a small local magnetization. 
When $q>1$, activation is concentrated in nodes with high local active link density, and state imitation strengthens local magnetization, expanding state clusters in space. 
Concurrently, rewiring failures are concentrated at domain interfaces, leading to the spatial fragmentation of opinions. 
Conversely, when $q<1$, nodes are easily activated even by a small local active link density. 
Consequently, the increase in local magnetization caused by imitation cannot be sustained. 
Meanwhile, the accumulation of rewiring failures eliminates active links, leaving only the topological fragmentation of opinions.

Beyond the numerical results, we provided analytical approaches for the non-conserved properties and the absorption time $T_{\rm abs}$. The evolution of the mean degree $\langle{k}\rangle$ is described by a stretched exponential function, where the stretching exponent $\beta$ depends on $(p,q)$. Furthermore, by treating absorption as a finite-time first-passage process, we obtained the terminal mean degree and mean absorption time, demonstrating reasonable agreement with the numerical simulation results.

Our spatially constrained CNVM can be extended in various directions. While we introduced the RGG, our framework is applicable to any spatially embedded graph~\cite{barthelemySpatialNetworks2011}. In the present model, the mean degree does not increase because rewiring only redirects existing links; however, one could also consider an alternative rule in which active nodes create additional links. Our model and the degree-preserving CNVM represent two limiting cases: interactions restricted by spatial constraints and interactions without spatial constraints, respectively. In reality, however, individuals can be influenced not only by offline contacts in the real space but also by online contacts via social media. This interplay could be captured by a multiplex network~\cite{kivelaMultilayerNetworks2014} combining spatial and non-spatial interaction layers.

\begin{acknowledgments}
This study was supported by the Basic Science Research Program through the National Research Foundation of Korea (RS-2025-00514776).
\end{acknowledgments}

\appendix

\section{\label{app:k_means}Empirical regime classification by k-means clustering}
\setcounter{figure}{0}
\setcounter{table}{0}
\renewcommand{\thefigure}{A\arabic{figure}}
\renewcommand{\thetable}{A\arabic{table}}

We applied $k$-means clustering to the standardized feature vector defined in Eq.~\eqref{eq:feature_vector} of the main text.
Each component was standardized to zero mean and unit variance across the full parameter grid before clustering.
To choose the number of clusters, we scanned $k=3,\ldots,8$ and computed the silhouette score for each $k$. 
Although the inertia trivially decreases monotonically with increasing $k$, the silhouette score is maximal at $k=5$, with a score of $0.521$ and an inertia of $729.8$. The improvement over $k=4$ is modest ($0.516$ to $0.521$), so we use $k=5$ as an empirical partition rather than as evidence for a sharply determined number of physical phases. 

\begin{table}[h]
\caption{\label{tab:kmeans_metrics} Silhouette scores and inertia for different numbers of clusters $k$. The maximum silhouette score is observed at $k=5$, although the difference from $k=4$ is small.}
\begin{ruledtabular}
\begin{tabular}{ccc}
$k$ & Silhouette score & Inertia \\
\colrule
3 & 0.459 & 1789.3 \\
4 & 0.516 & 1148.9 \\
5 & 0.521 & 729.8 \\
6 & 0.492 & 606.3 \\
7 & 0.502 & 490.2 \\
8 & 0.502 & 414.5 \\
\end{tabular}
\end{ruledtabular}
\end{table}

\section{\label{app:size_robustness}System size robustness of final state observables}
\setcounter{figure}{0}
\setcounter{table}{0}
\renewcommand{\thefigure}{B\arabic{figure}}
\renewcommand{\thetable}{B\arabic{table}}

\begin{figure}[h]
    \centering
    \includegraphics[width=\columnwidth]{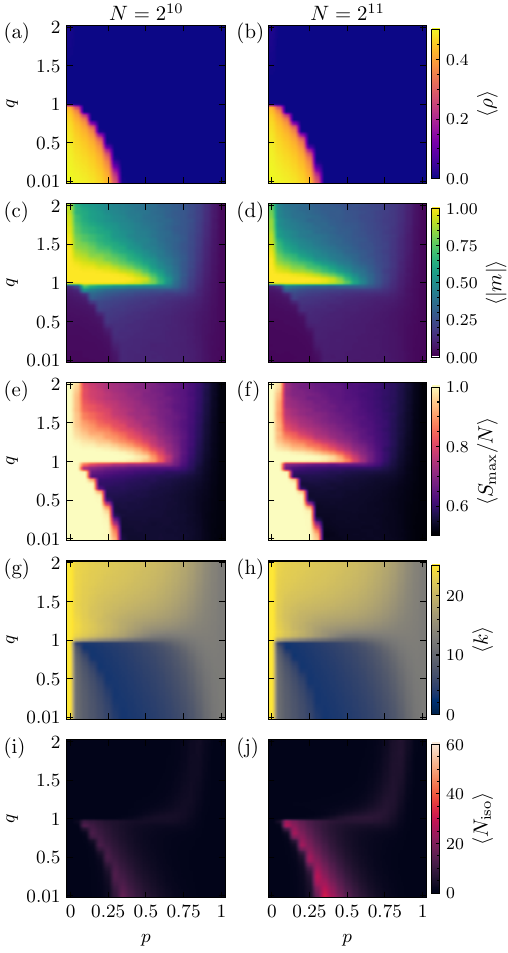}
    \caption{System size comparison of the final state observables used for the empirical regime map at $N=2^{10}$ and $N=2^{11}$. The update rule, observation time, and parameter grid are the same as in Fig.~\ref{fig:summary_six_panel}.}
    \label{fig:app_size_robustness}
\end{figure}

We checked whether the empirical regime structure depends sensitively on system size by comparing the final state observables for $N=2^{10}$ and $N=2^{11}$.
Fig.~\ref{fig:app_size_robustness} shows five observables using the same update rule, observation time, and parameter grid as in the main analysis. The main qualitative structures are already visible at these smaller sizes: the low-$p$, low-$q$ active regime, the high-magnetization branch near $q\simeq1$, the largest-component structure, and the regions of degree loss and isolated node formation. Thus the empirical regimes discussed in the main text are not artifacts of a single system size.

\section{\label{app:P4_finite_rewiring}Long-time test of the consensus regime (P4)}
\setcounter{figure}{0}
\setcounter{table}{0}
\renewcommand{\thefigure}{C\arabic{figure}}
\renewcommand{\thetable}{C\arabic{table}}

This appendix supports the finite-size interpretation of P4 used in the main text. We test whether the high magnetization regime remains globally ordered when finite rewiring is allowed over a longer observation time.

\begin{figure}[htbp]
    \centering
    \includegraphics[width=\columnwidth]{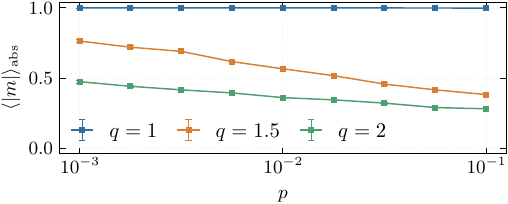}
    \caption{
    Long-time tests of the consensus-like regime.
    Conditional mean magnetization of absorbed runs,
    $\langle |m| \rangle_{\rm abs}$, as a function of rewiring probability $p$ for $q=1$, $1.5$, and $2$.
    These simulations use an extended cutoff $T_{\max}=2^{19}$ MCS.
    The absorption probability is $P_{\rm abs}=1$ for all shown points except $q=2$, $p=0.001$, where $P_{\rm abs}=0.992$. Error bars represent the standard error.}
    \label{fig:app_P4}
\end{figure}

Fig.~\ref{fig:app_P4} shows additional simulations with an extended cutoff $T_{\max}=2^{19}$ MCS. The plotted magnetization is the conditional mean over absorbed runs, $\langle|m|\rangle_{\rm abs}$. At this cutoff, absorption is nearly complete for all shown parameter points: $P_{\rm abs}=1$ except at $q=2$, $p=0.001$, where $P_{\rm abs}=0.992$. For $q=1$, the absorbed-run magnetization remains near one over the small-$p$ range shown. For $q=1.5$ and $q=2$, weak rewiring already suppresses $\langle|m|\rangle_{\rm abs}$.

These long-time tests support the interpretation that the high-$q$ part of the consensus-like regime corresponds to fragmented absorbing configurations with reduced global magnetization, rather than to a stable thermodynamic consensus phase.

\section{\label{app:degree_k}Low-degree tail for \texorpdfstring{$q=2$}{q=2}}
\setcounter{figure}{0}
\setcounter{table}{0}
\renewcommand{\thefigure}{D\arabic{figure}}
\renewcommand{\thetable}{D\arabic{table}}

\begin{figure}[htbp]
    \centering
    \includegraphics[width=\linewidth]{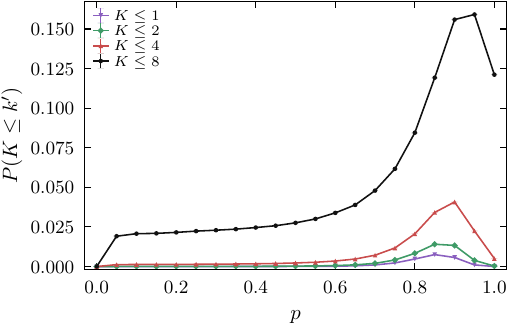}
    \caption{
        Cumulative low-degree fraction $P(K \le k')$ at $q=2$ for $k'=1,2,4,8$. Isolated nodes with $K=0$ are included. The low-degree tail is largest at large $p<1$, consistent with the peak position of the isolated node count. Error bars represent the standard error.
    }
    \label{fig:supp_low_degree_tail_q2}
\end{figure}

To further examine why the isolated node count for $q=2$ peaks at a large value below $p=1$, we analyze the final degree distribution. Fig.~\ref{fig:supp_low_degree_tail_q2} shows the cumulative fraction of nodes with degree $K \le k'$ for $k'=1,2,4,8$, including isolated nodes. The low-degree tail is enhanced around the same high-$p$ region where $\langle N_{\rm iso}\rangle$ peaks, rather than at $p=1$, where the total failed rewiring count is largest. This supports the interpretation that isolated nodes are controlled by local degree loss and the removal of the remaining few links, not simply by the total number of failed rewiring events.

\section{\label{app:abs_time}Absorption time at representative point}
\setcounter{figure}{0}
\setcounter{table}{0}
\renewcommand{\thefigure}{E\arabic{figure}}
\renewcommand{\thetable}{E\arabic{table}}
As an additional check, we directly measured the absorption time at a representative point in the finite-time active regime, $p=q=0.1$. Fig.~\ref{fig:app_abs_time} shows the mean absorption time in MCS units as a function of system size. Over the accessible range of $N$, $\langle T_{\rm abs}\rangle$ remains of the same order and does not show systematic growth with $N$. All sampled runs reached an absorbing state within the extended simulation time. 

\begin{figure}[h]
    \centering
    \includegraphics[width=\columnwidth]{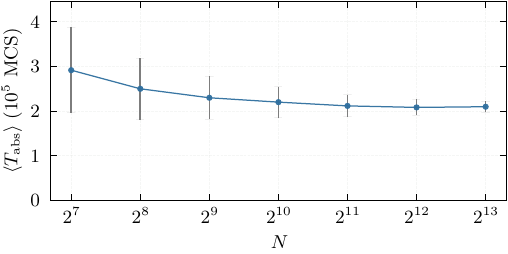}
    \caption{Symbols show the mean absorption time
    $\langle T_{\rm abs}\rangle$ in MCS units, and error bars indicate the
    standard deviation over runs. The weak size dependence indicates that the absorption time remains of the same order in MCS units over the accessible system sizes.}
    \label{fig:app_abs_time}
\end{figure}

\section{\label{app:eff_fss}Finite-time effective scaling of the survival probability}
\setcounter{figure}{0}
\setcounter{table}{0}
\renewcommand{\thefigure}{F\arabic{figure}}
\renewcommand{\thetable}{F\arabic{table}}
In the main text, we identify P5 as a finite-time active regime, or equivalently as a regime where a substantial fraction of runs remain nonabsorbed up to the observation time $T_{\max}$.
Near the P5--P1 boundary, the ensemble is naturally divided into two classes of trajectories.
Some runs reach an absorbing state within the observation time, so that $\rho=0$.
Other runs survive until $T_{\max}$ with a nonzero active link density, $\rho(T_{\max})>0$.

Because of this mixture, the ensemble-averaged active link density combines two effects:
the fraction of runs that survive and the typical active link density within those surviving runs.
Schematically,
\begin{equation}
\langle \rho(T) \rangle
\simeq
P_{\rm surv}(T)
\langle \rho(T)\mid \rho(T)>0\rangle ,
\end{equation}
where
\begin{equation}
P_{\rm surv}(T)
=
{\rm Prob}[\rho(T)>0]
\end{equation}
is the probability that a run has not reached an absorbing state by time $T$.
Thus, $P_{\rm surv}$ is the more direct observable for quantifying the finite-time active character of P5. This survival-conditioned viewpoint is closely related to the quasi-stationary treatment of stochastic processes with absorbing states~\cite{dickmanQuasistationaryDistributionsStochastic2002}.

\begin{figure}[h]
\centering
\includegraphics[width=\columnwidth]{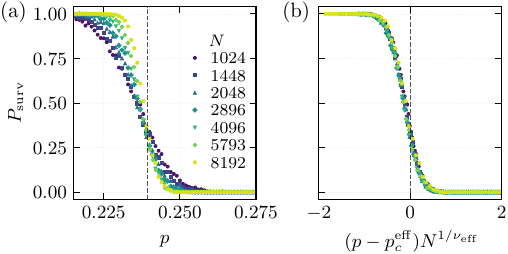}
\caption{
Effective finite-size scaling of the finite-time survival probability.
(a) Survival probability $P_{\rm surv}(p,N;T_{\max})$ at $q=0.5$ and
$T_{\max}=2^{15}$ MCS for different system sizes.
The vertical dashed line indicates the fitted effective threshold $p_c^{\rm eff}$.
(b)  Effective collapse at fixed $T_{\max}$ with $p_c^{\rm eff}\simeq 0.239$ and $1/\nu_{\rm eff}\simeq 0.481$.}
\label{fig:finite_time_active}
\end{figure}

To characterize the finite-time boundary between the active and absorbing regimes, we analyze $P_{\rm surv}(p,N;T)$ at $q=0.5$.
At fixed observation time, the survival probability can be organized by an effective finite-size scaling form,
\begin{equation}
P_{\rm surv}(p,N;T)
=
F\left[
(p-p_c^{\rm eff}(T))N^{1/\nu_{\rm eff}}
\right].
\end{equation}
For the collapse shown in Fig.~\ref{fig:finite_time_active}, we use $T=T_{\max}=2^{15}$ MCS and obtain
\begin{equation}
p_c^{\rm eff}\simeq0.239,
\qquad
1/\nu_{\rm eff}\simeq0.481 .
\end{equation}

Therefore, the scaling collapse should be interpreted as an effective description at fixed observation time, rather than as a determination of an asymptotic active phase transition.

\section{\label{app:absorbing_closure}Details of the absorption-route closure}
\setcounter{figure}{0}
\setcounter{table}{0}
\renewcommand{\thefigure}{G\arabic{figure}}
\renewcommand{\thetable}{G\arabic{table}}
The reduced stopping process model uses three effective absorption routes, indexed by $i\in\{L,M,H\}$. Each route is specified by a route weight $\alpha_i(p,q)$, a cumulative first-passage law $F_i(T;p,q)$, and a terminal degree function $d_i(p)$. In the fits reported in Sec.~\ref{sec:absorption_model}, the stage numbers are fixed as
\begin{equation}
(n_L,n_M,n_H)=(1,4,16).
\end{equation}
The continuous coefficients entering the route rates, route weights, and terminal degree functions are estimated from the simulation data.
The labels $L$, $M$, and $H$ denote terminal degree branches, while the fixed Erlang orders provide a reduced basis for the first-passage laws. They are model-order choices, not microscopic stage counts.
The absorption time along route $i$ is modeled as an Erlang random variable,
\begin{equation}
T_i\sim {\rm Erlang}(n_i,\lambda_i).
\end{equation}
Equivalently, route $i$ is represented by $n_i$ effective transient stages traversed sequentially at rate $\lambda_i(p,q)$. This defines a restricted phase-type basis in which absorption occurs after a finite sequence of Markovian stages. Phase-type laws are standard absorption time distributions of finite Markov processes, and the Erlang law corresponds to the special case of identical sequential stages.
The cumulative absorption probability along route $i$ is
\begin{equation}
F_i(T;p,q)=\Pr(T_i\le T)=P\!\left(n_i,\lambda_i(p,q)T\right),
\label{eq:app_Fi_gamma}
\end{equation}
where $P(n,x)$ is the regularized lower incomplete gamma function. Since $n_i$ is an integer, Eq.~\eqref{eq:app_Fi_gamma} can be written as
\begin{equation}
F_i(T;p,q)=1-e^{-\lambda_i(p,q)T}\sum_{r=0}^{n_i-1}\frac{\left[\lambda_i(p,q)T\right]^r}{r!}.
\label{eq:app_Fi_closed}
\end{equation}
The corresponding density is
\begin{equation}
\partial_tF_i(t;p,q)=\frac{\lambda_i(p,q)^{n_i}t^{n_i-1}e^{-\lambda_i(p,q)t}}{(n_i-1)!}.
\label{eq:app_fi}
\end{equation}

The route rates are parameterized as
\begin{equation}
\lambda_i(p,q)=\exp\left(\beta_{i0}+\beta_{ip}p+\beta_{iq}q\right),
\label{eq:app_lambda}
\end{equation}
and the route weights as
\begin{equation}
\alpha_i(p,q)=\frac{\exp(a_{i0}+a_{ip}p+a_{iq}q)}{\sum_{j\in\{L,M,H\}}\exp(a_{j0}+a_{jp}p+a_{jq}q)}.
\label{eq:app_alpha}
\end{equation}
This softmax form enforces nonnegative route weights and $\sum_i\alpha_i(p,q)=1$.

\begin{figure}[htbp]
\centering
\includegraphics[width=\columnwidth]{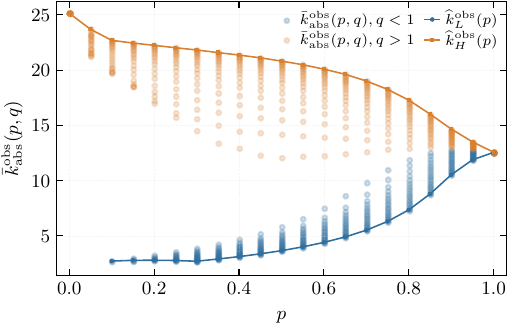}
\caption{
Empirical terminal degree envelopes.
The low-$q$ envelope $\hat{k}_L^{\rm obs}(p)$ captures the degree-eroding branch, while the high-$q$ envelope $\hat{k}_H^{\rm obs}(p)$ captures the degree-preserving branch.
The fitted endpoint functions define $d_L(p)$ and $d_H(p)$ used in the absorption-route closure.
}
\label{fig:terminal_degree_envelopes}
\end{figure}

The terminal degree functions are constructed from empirical envelopes of the terminal degree in absorbed realizations. Let
\begin{equation}
\bar{k}_{\rm abs}^{\rm obs}(p,q)=\mathbb{E}\left[k_{\rm term}^{(a)}(p,q)\,\middle|\,T_{\rm abs}^{(a)}\le T_{\max}\right]
\end{equation}
denote the observed terminal mean degree conditional on finite-time absorption. The low-degree and high-degree envelopes are estimated as
\begin{equation}
\hat{k}_L^{\rm obs}(p)=\min_{q<1}\bar{k}_{\rm abs}^{\rm obs}(p,q),\qquad\hat{k}_H^{\rm obs}(p)=\max_{q>1}\bar{k}_{\rm abs}^{\rm obs}(p,q),
\label{eq:app_observed_envelopes}
\end{equation}
where the extrema are taken over parameter cells with absorbed realizations. These envelopes provide empirical endpoint constraints for the terminal degree functions and are shown in Fig.~\ref{fig:terminal_degree_envelopes}.

The lower envelope corresponds to the most strongly degree-eroding absorbed states observed on the sublinear-response side, whereas the upper envelope represents the most degree-preserving absorbing states observed on the superlinear-response side. The two envelopes are used as endpoint constraints for the terminal degree functions. We anchor the endpoint functions to three degree levels: the residual low-degree level $k_{\min}$, the half-degree level $k_{\rm half}=k_0/2$, and the initial degree level $k_0$. The half-degree level follows from the initially random state assignment. On average, half of the initial neighbors of a node have the opposite state. Removing these discordant links in a rewiring-dominated absorbing branch gives the reference scale $k_0/2$.

We use the envelope parameterization
\begin{equation}
\frac{\hat{k}_L^{\rm obs}(p)-k_{\min}}{k_{\rm half}-k_{\min}}\simeq p^{\nu_L},\qquad\frac{\hat{k}_H^{\rm obs}(p)-k_{\rm half}}{k_0-k_{\rm half}}\simeq(1-p)^{\nu_H}.
\label{eq:app_rescaled_envelopes}
\end{equation}
This gives
\begin{equation}
d_L(p)=k_{\min}+(k_{\rm half}-k_{\min})p^{\nu_L},\qquad\nu_L>0,
\label{eq:app_dL}
\end{equation}
and
\begin{equation}
d_H(p)=k_{\rm half}+(k_0-k_{\rm half})(1-p)^{\nu_H},\qquad\nu_H>0.
\label{eq:app_dH}
\end{equation}
The mixed terminal degree function is constrained by
\begin{equation}
d_M(p)=\omega_M d_L(p)+(1-\omega_M)d_H(p),\qquad0\le\omega_M\le1.
\label{eq:app_dM}
\end{equation}
Thus $d_L$ describes the degree-eroding absorbing branch, $d_H$ describes the degree-preserving absorbing branch, and $d_M$ gives a constrained intermediate assignment.

The fitted parameters are the rate coefficients
\begin{equation}
\{\beta_{i0},\beta_{ip},\beta_{iq}\}_{i\in\{L,M,H\}},
\end{equation}
the softmax coefficients
\begin{equation}
\{a_{i0},a_{ip},a_{iq}\}_{i\in\{L,M,H\}},
\end{equation}
and the terminal degree parameters
\begin{equation}
\{\nu_L,\nu_H,\omega_M,k_{\min}\}.
\end{equation}
The initial degree scale $k_0$ is measured from the initial graph ensemble used in the corresponding data set, and $k_{\rm half}=k_0/2$ follows from the unbiased initial state composition. The stage numbers $(1,4,16)$ are treated as fixed model-order choices.

For notation, define the finite window route moment
\begin{equation}
{\cal M}_i[\phi](p,q)=\int_0^{T_{\max}}\phi_i(t)\,\partial_tF_i(t;p,q)\,dt .
\label{eq:app_route_moment}
\end{equation}
Substitution into the finite-time stopping law gives
\begin{equation}
{\cal Z}(p,q)=\sum_{i\in\{L,M,H\}}\alpha_i(p,q){\cal M}_i[1](p,q).
\label{eq:app_Z}
\end{equation}
The conditional mean absorption time is
\begin{equation}
\tau_{\rm abs}(p,q)=\frac{\sum_i\alpha_i(p,q){\cal M}_i[t](p,q)}{{\cal Z}(p,q)}.
\label{eq:app_tau_moment}
\end{equation}
For the Erlang block,
\begin{equation}
{\cal M}_i[t](p,q)=\frac{n_i}{\lambda_i(p,q)}P\!\left(n_i+1,\lambda_i(p,q)T_{\max}\right).
\label{eq:app_erlang_first_moment}
\end{equation}
Therefore,
\begin{equation}
\tau_{\rm abs}(p,q)=\frac{\sum_i\alpha_i(p,q)\frac{n_i}{\lambda_i(p,q)}P\!\left(n_i+1,\lambda_i(p,q)T_{\max}\right)}{{\cal Z}(p,q)}.
\label{eq:app_tau_closed}
\end{equation}
Finally, the terminal mean degree conditioned on finite-time absorption is
\begin{equation}
k_{\rm abs}(p,q)=\frac{\sum_i\alpha_i(p,q){\cal M}_i[d_i](p,q)}{{\cal Z}(p,q)}.
\label{eq:app_k_closed}
\end{equation}
Equations~\eqref{eq:app_Z}, \eqref{eq:app_tau_closed}, and \eqref{eq:app_k_closed} are the closed forms used to compare the reduced stopping process closure with the simulation data.

\bibliographystyle{apsrev4-2}
\bibliography{bib}
\end{document}